\documentclass[aip,apl,reprint,a4paper,floatfix,amsmath,amssymb,amsfonts,noshowpacs,citeautoscript]{revtex4-2}
\usepackage{newtxtext,newtxmath}
\usepackage[T1]{fontenc}
\usepackage{graphicx}
\usepackage[dvipsnames]{xcolor}
\usepackage{soul} 
\usepackage{floatrow}
\usepackage{siunitx}
\usepackage[
,textwidth=17.5cm
,textheight=23.5cm
,verbose
,pdftex
]{geometry}
\usepackage{xspace}
\usepackage{svg}
\usepackage[pdftex]{hyperref}
\hypersetup{
  pdftitle={Improved operating voltage in InGaN-capped AlGaN-based DUV LEDs on bulk AlN substrates},
	colorlinks,
	citecolor=blue,
	linkcolor=blue
	}

\makeatletter
\newcommand*{\refig}[2]{\hyperref[#1]{\ref*{#1}(#2)}}
\makeatother

\DeclareMathAlphabet{\mathsf}{OT1}{\sfdefault}{m}{n}
\SetMathAlphabet{\mathsf}{bold}{OT1}{\sfdefault}{b}{n}

\begin{document}

\preprint{AIP/123-QED}

\title[Sample title]{Improved operating voltage in InGaN-capped AlGaN-based DUV LEDs on bulk AlN substrates}
\author{Hsin-Wei S. Huang}
\email{hh494@cornell.edu}
\affiliation{\hbox{Department of Electrical and Computer Engineering, Cornell University, Ithaca, New York 14853, USA}}

\author{Shivali Agrawal}
\email{sa2368@cornell.edu}
\affiliation{\hbox{Department of Chemical and Biomolecular Engineering, Cornell University, Ithaca, New York 14853, USA}}

\author{Debaditya Bhattacharya}
\affiliation{\hbox{Department of Electrical and Computer Engineering, Cornell University, Ithaca, New York 14853, USA}}

\author{Vladimir Protasenko}
\affiliation{\hbox{Department of Electrical and Computer Engineering, Cornell University, Ithaca, New York 14853, USA}}

\author{Huili Grace Xing}
\affiliation{\hbox{Department of Electrical and Computer Engineering, Cornell University, Ithaca, New York 14853, USA}}
\affiliation{\hbox{Department of Materials Science and Engineering, Cornell University, Ithaca, New York 14853, USA}}
\affiliation{\hbox{Kavli Institute at Cornell for Nanoscale Science, Cornell University, Ithaca, New York 14853, USA}}

\author{Debdeep Jena}
\email{djena@cornell.edu}
\affiliation{\hbox{Department of Electrical and Computer Engineering, Cornell University, Ithaca, New York 14853, USA}}
\affiliation{\hbox{Department of Materials Science and Engineering, Cornell University, Ithaca, New York 14853, USA}}
\affiliation{\hbox{Kavli Institute at Cornell for Nanoscale Science, Cornell University, Ithaca, New York 14853, USA}}
\begin{abstract}
Better wall plug efficiency of deep-ultraviolet light emitting diodes (DUV-LEDs) requires simultaneous low resistivity p-type and n-type contacts, which is a challenging problem. In this study, the co-optimization of p-InGaN and n-AlGaN contacts for DUV LEDs are investigated. We find that using a thin In$_{0.07}$Ga$_{0.93}$N cap is effective in achieving ohmic p-contacts with specific contact resistivity of 3.10$\times$10$^{-5}$ $\Omega$cm$^2$. Upon monolithic integration of p- and n-contacts for DUV LEDs, we find that the high temperature annealing of 800$^{\circ}$C required for the formation of low resistance contacts to n-AlGaN severely degrades the p-InGaN layer, thereby reducing the hole concentration and increasing the specific contact resistivity to 9.72$\times$10$^{-4}$ $\Omega$cm$^2$. Depositing a SiO$_{2}$ cap by plasma-enhanced atomic layer deposition (PE-ALD) prior to high temperature n-contact annealing restores the low p-contact resistivity, enabling simultaneous low-resistance p- and n-contacts. DUV-LEDs emitting at 268 nm fabricated with the SiO$_{2}$ capping technique exhibit a 3.5 V reduction in operating voltage at a current level of 400 A/cm$^2$ and a decrease in differential ON-resistance from 6.4 m$\Omega$cm$^2$ to 4.5 m$\Omega$cm$^2$. This study highlights a scalable route to high-performance, high-Al-content bipolar AlGaN devices.  
\end{abstract}

\maketitle
AlGaN-based light-emitting diodes (LEDs) in the UVC range (200 nm-280 nm) are promising for applications in disinfection, sensing, and UV communications\cite{zhang2019271,lang2025progress}, offering compact, mercury-free alternatives to traditional UV sources. However, their wall plug efficiency (WPE) remains below 15$\%$\cite{lang2025progress}, far from the $\sim$30$\%$ required to compete with mercury lamps. A key limitation is the difficulty of forming low-resistance ohmic contacts to high-Al-content Al$_{x}$Ga$_{1-x}$N (x > 0.7)\cite{kneissl2016iii}, where both n- and p-type contacts suffer from intrinsic material challenges. As the Al composition increases, the electron affinity of AlGaN decreases from 3.18 eV for GaN to 1.01 eV for AlN, and suitable low work-function metals are limited, necessitating very high Si doping and annealing temperatures > 800$^{\circ}$C to achieve low-resistance n-contacts \cite{sulmoni2020electrical,taniyasu2002intentional,nagata2017reduction}. Similarly, p-contacts to Al$_{x}$Ga$_{1-x}$N (x > 0.4) suffer from large contact resistances due to the high ionization energy of Mg acceptors and the scarcity of large work-function metals \cite{bai2012effect, greco2016ohmic}. 

Despite numerous efforts on achieving good ohmic contact on both p-GaN and n-AlGaN surfaces, the operating voltages of DUV LEDs and high Al composition AlGaN p-n diodes remain high \cite{agrawal2024ultrawide,hao2017improved}. One of the critical problems lies in the large difference between the annealing temperatures required for ohmic contact formation on p-GaN and n-AlGaN surfaces\cite{hao2017improved}. The challenge in maintaining simultaneous low contact resistance for both p-type and n-type contacts during the monolithic integration of DUV LEDs and laser diodes on a single wafer leads to elevated operating voltages, increased Joule heating, and degraded device performance, particularly at high current densities. \cite{zhang2022key} 

In this study, we reduce the operating voltage of the DUV LED by first showing that employing a p-InGaN capping layer on p-GaN exhibits ohmic contact behavior and reduces the p-contact specific contact resistivity by approximately three orders of magnitude compared to p-GaN alone. We next observe that the > 800 °C anneal required for low-resistance n-contacts in LED fabrication severely degrades the p-InGaN layer, increasing the specific contact resistivity by over an order of magnitude. To address this, we introduce a thin SiO$_{2}$ cap on p-InGaN before n-contact annealing, preserving low p-contact resistance along with a low n-contact resistance as well. This work is helpful in reducing the operating voltage and series resistance of high-Al-content AlGaN-based bipolar devices.

\begin{figure*}
\includegraphics[width=\textwidth]{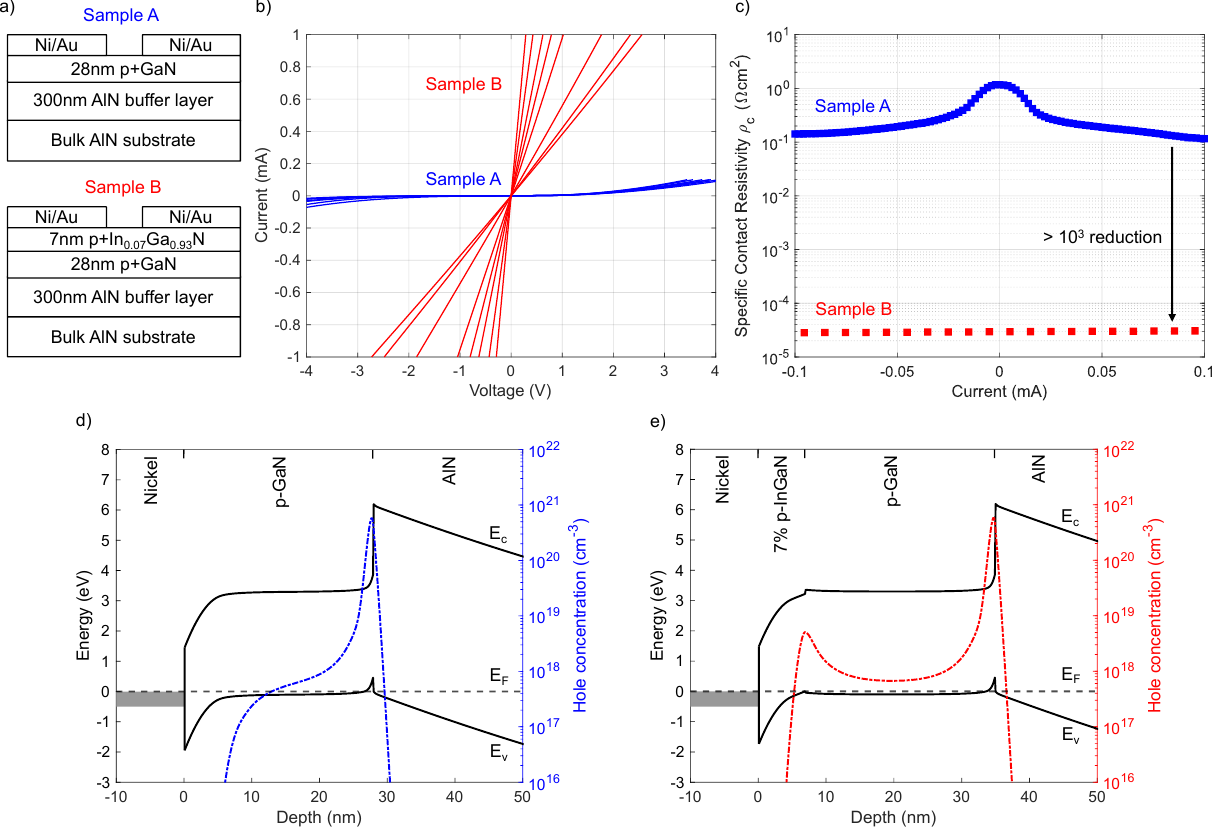}
\caption{\label{fig:1}(a) Schematic of the p-GaN contact samples without (sample A) and with (sample B) a p-InGaN cap. (b) CTLM-IV curves from the two samples. (c) Specific contact resistivity vs current level for the data shown in (b). The contact resistance of sample B is $R$$_{\mathrm{c}}$ = 8.4~$\Omega$mm. 
(d)-(e) Energy band diagram simulations of sample A and sample B, respectively.}
\end{figure*}

Two structures were fabricated to evaluate the effect of lowering the p-type contact resistance using a p-InGaN capping layer. The samples used in this study were grown using a nitrogen plasma-assisted Veeco Gen10 molecular beam epitaxy (MBE) system on +c-plane single-crystal bulk AlN substrates. Sample A consisted of a 300 nm thick AlN buffer layer and a 28 nm thick Mg-doped GaN. Sample B had an identical structure, with the addition of a 7 nm thick Mg-doped In$_{0.07}$Ga$_{0.93}$N cap. The sample schematics are shown in Fig.~\ref{fig:1}~(a). The two samples first underwent standard solvent cleaning (acetone and isopropanol) followed by lithographic patterning using nLOF 2020 and exposure with a 200 DSW i-line wafer stepper. Native oxide removal was performed using HCl:DI and buffered oxide etch (BOE), each for 15 seconds, followed by nitrogen drying. Finally, p-type metal–semiconductor contacts were formed via electron beam evaporation of Ni/Au (15/20 nm) and rapid thermal annealed (RTA) at 450$^{\circ}$C for 30 seconds in an O$_{2}$ atmosphere.

To study the effects of the p-InGaN capping layer, the specific contact resistivity and sheet resistance were characterized via the circular transmission line method (CTLM). The CTLM-IV measurements are shown in Fig.~\ref{fig:1}~(b). The TLM patterns had inner contact radius of 20 \textmu m and spacings from 1 \textmu m to 20 \textmu m. The contact of sample A  exhibits a non-linear I–V characteristic, while that of sample B exhibits a linear I-V characteristic. 

Fig.~\ref{fig:1}~(c) shows the specific contact resistivity vs current, extracted using $R$ = $dV/dI$.  At the current level of 0.1 mA, sample A exhibited a specific contact resistivity of $\rho$$_{\mathrm{c}}$ = 1.15 $\times$10$^{-1}$ $\Omega$cm$^2$ and sheet resistance of $R$$_{\mathrm{sheet}}$ = 91.9 k$\Omega$/$\square$. In contrast, sample B yielded a significantly lower specific contact resistivity of $\rho$$_{\mathrm{c}}$ = 3.10$\times$10$^{-5}$ $\Omega$cm$^2$ and sheet resistance of $R$$_{\mathrm{sheet}}$ = 24.2 k$\Omega$/$\square$. The contact resistance of sample B is $R$$_{\mathrm{c}}$ = 8.4~$\Omega$mm, comparable to the highest quality p-type InGaN contacts reported in literature \cite{dill2025velocity,bader2019gan,kumakura2003ohmic}. Notably, the specific contact resistivity of sample B is reduced by over three orders of magnitude relative to sample A. 

\begin{figure*}
\includegraphics[width=\textwidth]{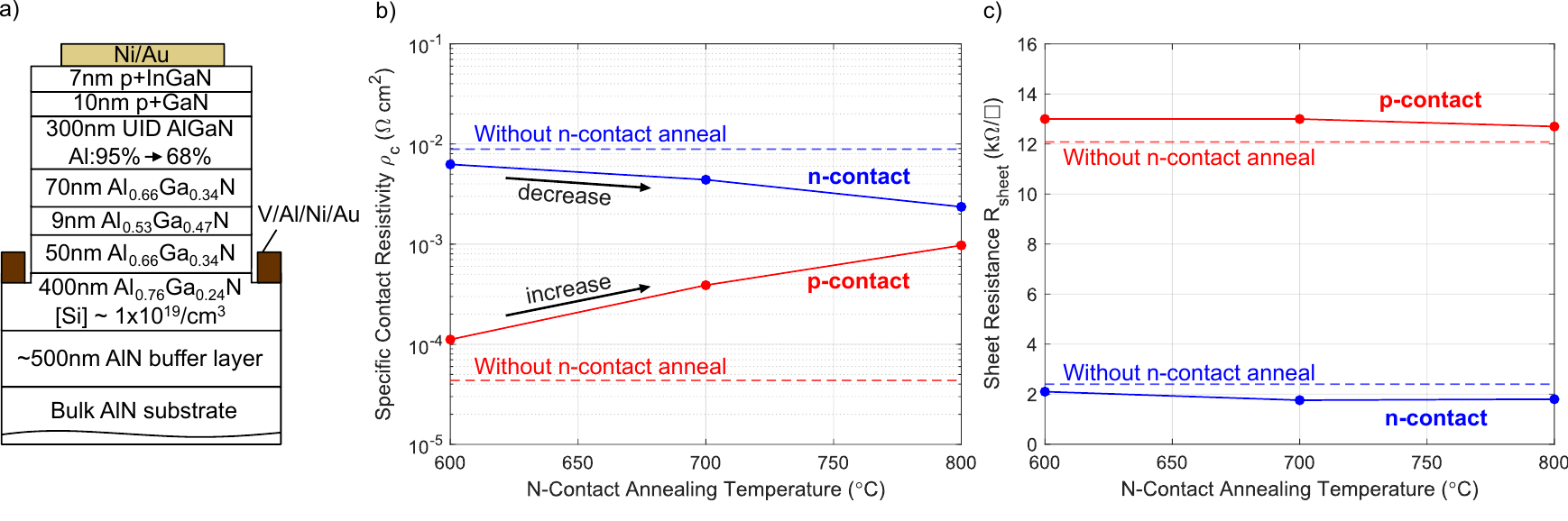}
\caption{\label{fig:2}(a) Heterostructure of the DUV LED samples used for this contact annealing temperature-dependent study. (b) Specific contact resistivity of n- and p-contact vs n-contact annealing temperature. P-contacts were subsequently annealed at 450 $^{\circ}$C. All resistance values were extracted at 1 mA from CTLM-IV measurement. (c) Sheet resistance vs n-contact annealing temperature.}
\end{figure*}

This substantial $\rho$$_{\mathrm{c}}$ reduction can be explained by the simulated energy band diagrams. These diagrams for samples A and B were computed using the self-consistent Schrödinger-Poisson solver, 1D Poisson \cite{snider1dpoisson}, and are presented in Fig.~\ref{fig:1}~(d) and (e), respectively.  In the simulation, nickel was used as the Schottky contact metal, with the theoretical barrier height  $q\phi_b$ calculated as \[q\phi_b = E_g-q(\phi_m-\chi_s),\] where the work function of Ni is $q\phi_m$ = 5.25 eV \cite{greco2016ohmic}, the bandgaps are $E_g$ = 3.4 eV for GaN\cite{alam2020bandgap} and 3.11 eV for In$_{0.07}$Ga$_{0.93}$N \cite{alam2020bandgap}, and the electron affinities are $\chi_s$ = 3.8 eV\cite{lin2012experimental} for GaN and $\chi_s$ = 3.86 eV for In$_{0.07}$Ga$_{0.93}$N \cite{lin2012experimental}, the latter obtained by linear interpolation using Vegard’s law. For both samples a uniform Mg doping concentration of 5$\times$10$^{19}/$cm$^{3}$ was assumed in the p-GaN and p-InGaN layers. The activation energy of Mg acceptors was taken as $\sim$180 meV for p-GaN \cite{zhao2017activation} and $\sim$140 meV for p-In$_{0.07}$Ga$_{0.93}$N \cite{zhao2017activation}.

A comparison of the energy band diagrams shows that employing the p-InGaN cap lowers the Schottky barrier height by 0.23~eV, allowing for enhanced thermionic emission of holes through the barrier. According to the thermionic emission (TE) and thermionic field emission model (TFE) \cite{sze2021physics}, lowering the Schottky barrier by 0.23~eV can provide three orders of reduction in the specific contact resistivity in p-GaN. Furthermore, the energy band diagrams show a narrowing of the tunneling barrier width in sample B. This is attributed to two key mechanisms: 1) the higher obtained hole concentration in the p-InGaN cap and 2) the strain-induced piezoelectric and spontaneous polarization fields that lead to band bending and result in a peak in the hole distribution at the pInGaN/pGaN interface. Together, these mechanisms contribute to the significant reduction in $\rho$$_{\mathrm{c}}$ observed with the use of a p-InGaN cap.  

The hole concentration within the p-type layers in sample A and B can be estimated from the measured sheet resistance using the relation
$R$$_{\mathrm{sheet}}$ = ($q$ $p$ $\mu$ $t$)$^{-1}$, where $q$ is the elementary charge 1.6$\times$10$^{-19}$ C, $p$ is the free hole concentration, $\mu$ is the hole mobility, and $t$ is the effective thickness of the conducting layer. Using the measured sheet resistances of $R$$_{\mathrm{sheet}}$ = 91.9 k$\Omega$/$\square$ and 24.2 k$\Omega$/$\square$ for samples A and B, respectively, layer thickness of $t$$_{\mathrm{pGaN}}$ = 28 nm for sample A and $t$$_{\mathrm{pInGaN/pGaN}}$ = 7 + 28 nm for sample B (accounting for parallel conduction through both layers), and assuming a hole mobility of $\mu$ = 10 cm$^{2}$V$^{-1}$s$^{-1}$ \cite{dai2021high}, the estimated hole concentrations are $p$ = 2.43$\times$10$^{18}$/cm$^3$ for sample A and 7.38$\times$10$^{18}$/cm$^3$ for sample  B. These estimated hole concentrations are higher than those induced by Mg-doping alone, which can be potentially ascribed to the two-dimensional hole gas (2DHG) at the InGaN/GaN and the GaN/AlN interface as predicted by the energy band diagram simulations\cite{chaudhuri2019polarization,chaudhuri2022very}. 

From a materials perspective, achieving hole concentrations exceeding 10$^{18}/$cm$^{3}$ is more feasible with p-InGaN than p-GaN \cite{kumakura2000activation}. First, Mg incorporation in InGaN alloy is significantly higher than in GaN. Lee et al. in ref.[\citenum{KevinLee_thesis}] reported that Mg-doped In$_{0.16}$Ga$_{0.84}$N incorporates 30$\times$ more Mg than GaN at the same growth temperature. Second, the high activation energy ($\sim$180 meV) of Mg acceptors in GaN strongly limits the free hole concentration at room temperature, with only 1-5$\%$ of incorporated Mg contributing to free carriers \cite{kumar2023growth,dai2021high}. Increasing the indium content in InGaN reduces this activation energy\cite{kumakura2000activation}, further boosting the hole concentration to above 10$^{18}/$cm$^{3}$ and reducing $\rho$$_{\mathrm{c}}$. It is worth noting that specific contact resistivities as low as $\rho$$_{\mathrm{c}}$ $\sim$ 10$^{-4}$-10$^{-5}$ $\Omega$ cm$^2$ have been reported for p-GaN contacts \cite{li2017design,ho1999low,cho2005characterization}. Achieving such low $\rho$$_{\mathrm{c}}$ in sample A would require optimizing the epitaxial growth and contact fabrication processes. In contrast, the incorporation of a p-InGaN capping layer offers a more direct and effective approach to reducing $\rho$$_{\mathrm{c}}$.   

\begin{figure*}
    \includegraphics[width=\textwidth]{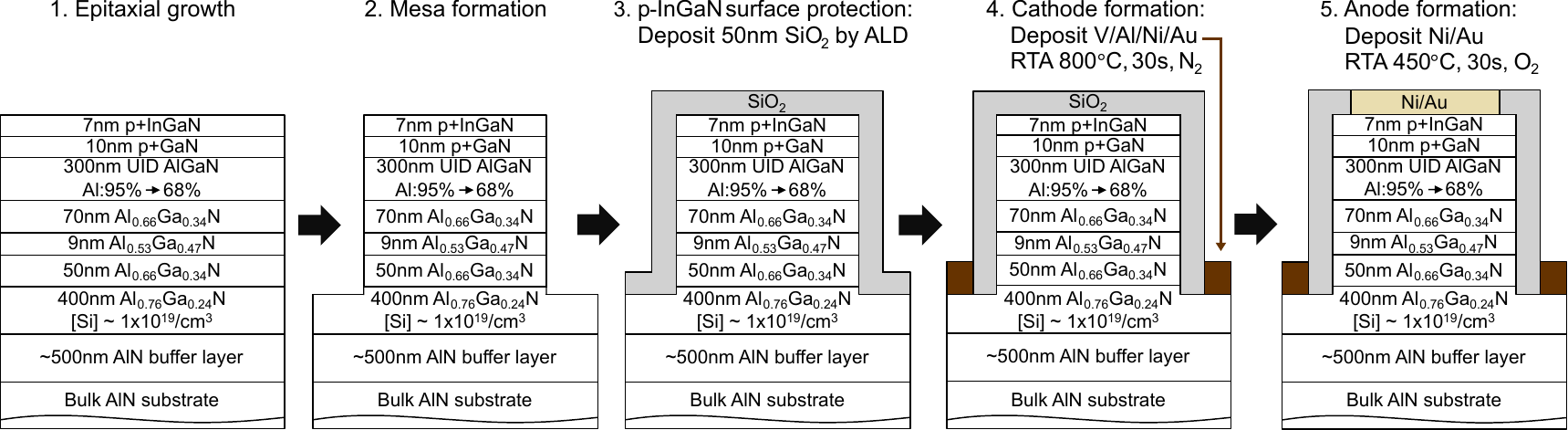}
    \caption{\label{fig:Figure 3}Schematic diagram illustrating the fabrication process of an LED with the SiO$_{2}$ capping technique. (a) Expitaxial growth. (b) Mesa formation by ICP-RIE. (c) Deposition of SiO$_{2}$ by ALD. (d) Deposition and annealing of n-type metal contacts (e) Deposition and annealing of p-type metal contacts.}
\end{figure*}

\begin{figure*}[htbp]
    \includegraphics[width=15 cm]{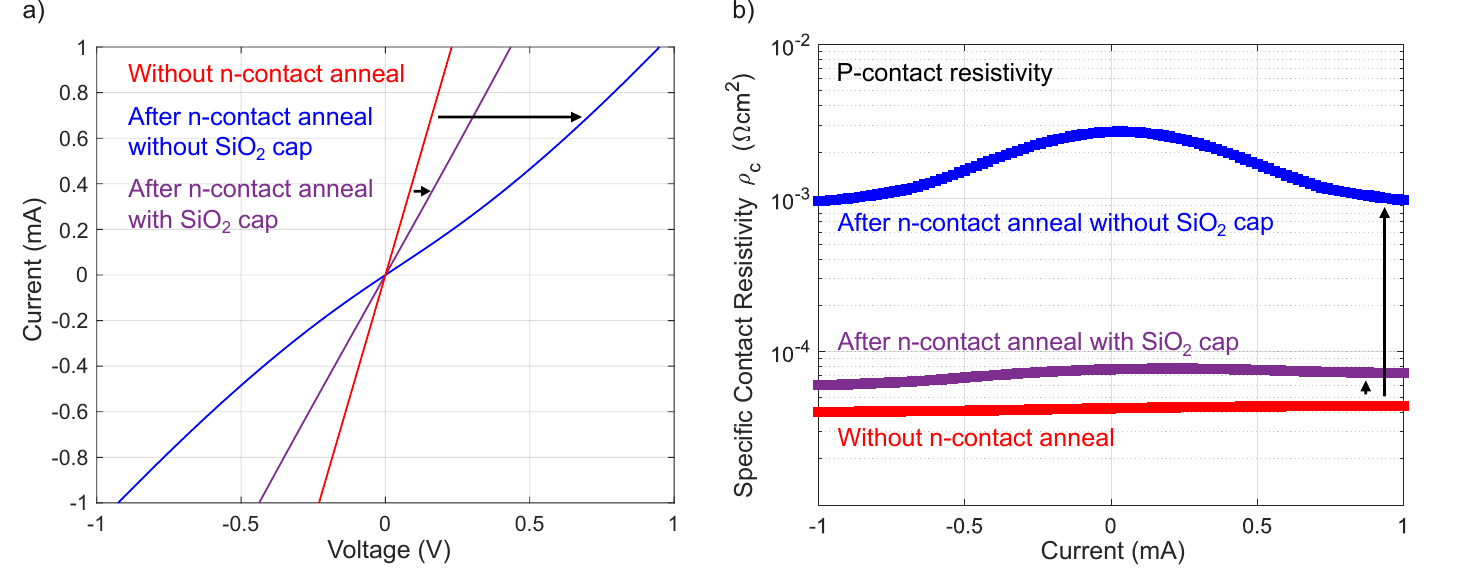}
    \caption{\label{fig:Figure 4}(a) CTLM-IV curves comparison of p-contact without undergoing n contact anneal, after undergoing n-contact anneal with SiO$_{2}$ cap, and after undergoing n-contact anneal without SiO$_{2}$ cap. IVs are plotted for 2 \textmu m spacing. (b) Specific contact resistivity vs current level for the data shown in (a).}
\end{figure*}

We next investigate the monolithic integration of p-InGaN and n-AlGaN contacts in DUV LED process. Specifically, we examine the impact of the high-temperature anneal required for low-resistance n-AlGaN contacts on the integrity and performance of the p-InGaN layer. Fig.~\ref{fig:2}~(a) shows the schematic of the DUV LED structure, which was grown on a 2-inch bulk AlN substrate by MBE. The heterostructure was grown along the [0001] crystal direction and consists of a 500 nm AlN nucleation layer, a 400 nm n-Al$_{0.76}$Ga$_{0.24}$N with donor doping of [Si]$\sim$ 1$\times$10$^{19}$/cm$^3$, a 50 nm  Al$_{0.66}$Ga$_{0.34}$N UID waveguiding layer, a 9 nm Al$_{0.53}$Ga$_{0.47}$N single quantum well, a 70 nm  Al$_{0.66}$Ga$_{0.34}$N UID waveguiding layer, a 300 nm distributed polarization doped (DPD) AlGaN layer graded from AlN to Al$_{0.65}$Ga$_{0.35}$N along the growth direction capable of providing $\sim$5.7$\times$10$^{17}$/ cm$^3$ mobile hole density without any impurity doping\cite{agrawal2024ultrawide}, and a Mg-doped contact layer with 10 nm p-GaN and capped with a 7 nm p-In$_{0.05}$Ga$_{0.95}$N layer. We prepared three samples of the size 1$\times$1 cm$^2$, all of which were diced from the same 2-inch wafer. The LED devices were fabricated by first exposing the n-cladding layer surface by inductively-coupled plasma reactive ion etching (ICP RIE) $\sim$50 nm into the nAl$_{0.76}$Ga$_{0.24}$N layer. Next, n-electrode metal stack V/Al/Ni/Au with thicknesses 20/80/40/100 nm  was deposited using electron-beam evaporation. The three samples were annealed for 30 s in N$_{2}$ ambient at different temperatures: 600$^{\circ}$C, 700$^{\circ}$C, and 800$^{\circ}$C. Identical p-contact Ni/Au metal stack with 15/20 nm thicknesses were subsequently deposited on the three samples on the p-In$_{0.05}$Ga$_{0.95}$N surface and annealed for 30 s at 450$^{\circ}$C in O$_{2}$ ambient. 

CTLM structures were fabricated on the p-InGaN surface and the etched n-AlGaN surfaces of the three samples. The specific contact resistivity of both n- and p-type contacts as a function of n-contact annealing temperature is shown in Fig.~\ref{fig:2}~(b), with all values extracted at the current level of 1 mA. Horizontal lines mark $\rho$$_{\mathrm{c}}$ without n-contact anneal.

\begin{figure*}
    \includegraphics[width=\textwidth]{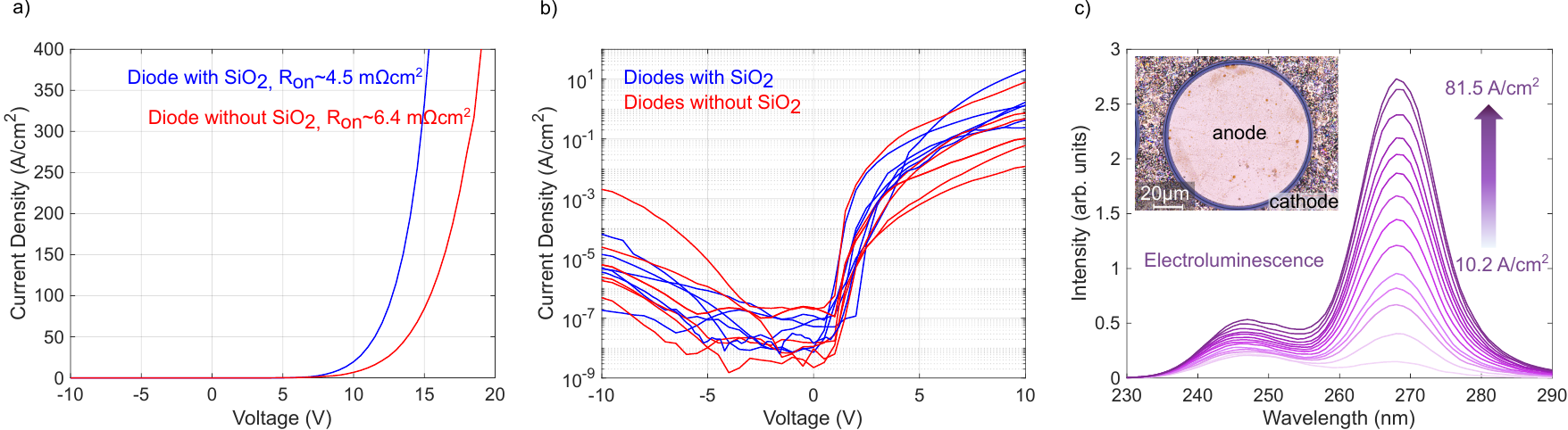}
    \caption{(a) Room temperature J-V characteristics of two LEDs, one with SiO$_{2}$ capping method and one without. The differential ON-resistance was extracted at 400 A/cm$^2$. (b) IV from batch test of LEDs with and without SiO$_{2}$ capping. (c) Room temperature electroluminescence of an LED with the SiO$_{2}$ capping method. Inset shows the microscopy image of a fabricated LED.}
    \label{fig:Figure 5}
\end{figure*}


For the n-contact, high annealing temperatures of 600$^{\circ}$C to 800$^{\circ}$C are required to lower $\rho$$_{\mathrm{c}}$, consistent with reports for high-composition-Al n-AlGaN contacts\cite{sulmoni2020electrical}. Several studies in literature have suggested that during the annealing of V/Al-based metal contacts with n-AlGaN, nitrogen is extracted from the semiconductor, forming a thin crystalline AlN at the metal–semiconductor interface from Al in the electrode stack\cite{lapeyrade2013electrical,haidet2017nonlinear}. Nitrogen vacancies act as donor states, increasing carrier concentration at the interface\cite{sulmoni2020electrical} and enhancing tunneling through the thin space charge region, thus reducing contact resistance. For high-Al composition AlGaN, studies have shown that higher annealing temperatures are required to drive the metal–nAlGaN reaction needed for low-resistance ohmic contacts\cite{lapeyrade2013electrical,van2005role,sulmoni2020electrical}, which agrees with the trend in the contact resistance data presented in Fig. 2 (b). The specific contact resistivity of our reported n-type contacts are higher than values reported in the literature, where $\rho$$_{\mathrm{c}}$ $\leq$ 5$\times$10$^{-4}$ $\Omega$cm$^2$ has been achieved for nAl$_{x}$Ga$_{1-x}$N (x > 0.7)\cite{sulmoni2020electrical,nagata2017reduction}. We attribute the elevated $\rho$$_{\mathrm{c}}$ in our devices to several possible factors: 1) surface damage and roughening caused by high-power RIE, and 2) the use of native AlN substrates rather than foreign sapphire substrates. While AlN substrates offer advantages such as lower threading dislocation and point defect densities\cite{kobayashi2024enhanced}, this reduced defect density may limit the effectiveness of defect-assisted tunneling mechanisms that otherwise help reduce $\rho$$_{\mathrm{c}}$ in devices grown on sapphire \cite{haidet2015conduction}. Further reduction in $\rho$$_{\mathrm{c}}$ may be achieved by increasing the Si doping concentration from 1$\times$10$^{19}/$cm$^{3}$ to 3$\times$10$^{19}/$cm$^{3}$, which is predicted to lower $\rho$$_{\mathrm{c}}$ to the $\sim$10$^{-5}$ $\Omega$cm$^2$ range for Al$_{0.75}$Ga$_{0.25}$N based on TFE model calculations. Additional strategies reported in the literature include surface treatment\cite{zhang2021improved,sarkar2017performance} and the incorporation of an intermediate interfacial layer \cite{nagata2017reduction}.        

In contrast to the n-type contacts, the specific contact resistivity of the p-type contacts increases with higher n-contact annealing temperatures. The lowest p-contact $\rho$$_{\mathrm{c}}$ = 4.37 $\times$10$^{-5}$ $\Omega$cm$^2$ was observed in the sample that did not undergo n-contact annealing. As the n-contact annealing temperature increased from no anneal to 800$^{\circ}$C, the p-contact rose super-linearly to 9.72 $\times$10$^{-4}$ $\Omega$cm$^2$, as shown in Fig.~\ref{fig:2}~(b). We attribute this degradation to thermal decomposition and nitrogen desorption from the p-InGaN surface—mechanisms previously reported for p-GaN contacts\cite{hao2017improved}. This process forms a high concentration of nitrogen vacancy defects near the sample surface that acts as compensating donors in the p-type layer\cite{ganchenkova2006nitrogen}, thereby reducing the hole concentration and increasing $\rho$$_{\mathrm{c}}$. 

These findings highlight a critical trade-off in monolithic DUV LED integration: while high-temperature annealing improves the n-contact resistivity, it simultaneously degrades the p-contact. To maintain low p-contact resistivity, the n-contact annealing temperature should ideally remain below 800$^{\circ}$C; however, this constraint may limit the performance of the n-contact. Similar p-GaN degradation has been reported\cite{hao2017improved, KevinLee_thesis}, which the authors mitigated by separately optimizing p- and n-contact anneals to minimize series resistance.

Additional insight can be drawn from the sheet resistance trends shown in Fig.~\ref{fig:2}~(c). While the p-contact sheet resistance increases from 12.17 k$\Omega$/$\square$ at no n-contact anneal to 13.1~k$\Omega$/$\square$ at 600$^{\circ}$C, it remains relatively constant up to 800$^{\circ}$C. This suggests that the temperature-induced degradation is localized primarily at the metal–semiconductor interface, rather than in the bulk of the p-InGaN layer. In contrast, the sheet resistance of the n-type contact decreases slightly with increasing annealing temperature, consistent with improved ohmic contact formation.

To mitigate thermal degradation and nitrogen evaporation of p-InGaN during high-temperature n-contact anneal, we deposited a 50 nm SiO$_{2}$ cap by PE-ALD, inspired by ref.  [\citenum{lee_annealing_2000}], which showed SiO$_{2}$ suppresses Mg out-diffusion during RTA of Mg-doped GaN. The full LED fabrication sequence incorporating the SiO$_{2}$ cap is shown in Fig.~\ref{fig:Figure 3}. To evaluate the effectiveness of this technique, TLM measurements were performed on three p-contact samples: one without n-contact annealing, one with 800$^{\circ}$C n-contact annealing without the SiO$_{2}$ cap, and one with the SiO$_{2}$ cap during the anneal. The results, shown in Fig.~\ref{fig:Figure 4} (a)-(b), indicate that the specific contact resistivity of the p-contact was significantly reduced from 9.72$\times$10$^{-4}$ $\Omega$cm$^2$ without the cap to 6.76$\times$10$^{-5}$ $\Omega$cm$^2$ with the cap, nearly restoring it to the value of 4.37$\times$10$^{-5}$ $\Omega$cm$^2$ observed in the un-annealed reference sample.  

In addition to lowering the p-contact resistance, the SiO$_{2}$ capping technique exhibited spatial uniformity across the sample. A 1$\times$1 cm$^2$ sample was partitioned into four regions, each containing a central CTLM structure. The average p-contact $\rho$$_{\mathrm{c}}$ across the four dies was 6.93 $\times$10$^{-5}$ $\Omega$cm$^2$ $\pm$ 0.38 $\times$10$^{-5}$ $\Omega$cm$^2$, indicating excellent uniformity and consistency. This method provides an effective strategy for mitigating thermal degradation of p-InGaN contacts. While academic journals seldom detail it explicitly, using dielectric (SiO$_{2}$, SiN$_{\mathrm{x}}$) or epitaxial cap layers (GaN, AlN) to protect device epi prior to contact deposition is a standard but under-reported processing step \cite{kumabe2024demonstration,schubert2023light}. However, there is currently no report of utilizing a dielectric capping layer to intentionally mitigate degradation of p-contacts in the DUV LED community. 

Two DUV LED samples were fabricated from the same epitaxial wafer, with a heterostructure similar to that shown in  Fig.~\ref{fig:Figure 3}. One sample used the SiO$_{2}$ capping technique and the other did not. Diodes were tested randomly from each sample, and Fig.~\ref{fig:Figure 5} (a) compares the room-temperature J–V characteristics of the best-performing diode from each group— selected based on the lowest ON-resistance at 10 V. At 400~A/cm$^2$, the SiO$_{2}$-capped diode showed reduced forward voltage to 15.2 V compared to 19.0 V for the uncapped diode, and lower ON-resistance of 4.5~m$\Omega$cm$^2$ versus 6.4~m$\Omega$cm$^2$. Batch testing, as shown in Fig.~\ref{fig:Figure 5} (b), confirmed improved performance. SiO$_{2}$-capped diodes averaged J$_{10\mathrm{V}}$ = 1.21$\pm$0.54 A/cm$^2$ while uncapped devices averaged J$_{10\mathrm{V}}$ = 0.34$\pm$0.22 A/cm$^2$. The variation within each group is attributed to fabrication non-uniformities. Most devices showed leakage current between 10$^{-7}$ A/cm$^2$ to 10$^{-5}$A/cm$^2$ at –10 V. 

Electroluminescence (EL) spectra, collected from the backside of a LED with SiO$_{2}$ capping at room temperature is shown in Fig.~\ref{fig:Figure 5} (c). These results collectively demonstrate that SiO$_{2}$ capping enhances AlGaN DUV LED performance by lowering series resistance and operating voltage.

In this work, we demonstrated that a p-InGaN capping layer enables low-resistance p-type contacts for AlGaN-based DUV LEDs on bulk AlN, reducing the specific contact resistivity to 3.10$\times$10$^{-5}$ $\Omega$cm$^2$—three orders lower than p-GaN alone—due to reduced Schottky barrier height, lower Mg activation energy, and polarization-induced band bending. However, high-temperature annealing for n-contacts degrades p-contact performance. Implementing a SiO$_{2}$ cap preserved the InGaN surface, recovering the specific contact resistivity to 6.76$\times$10$^{-5}$ $\Omega$cm$^2$. LEDs fabricated with this method showed reduced operating voltage and ON-resistance, alongside DUV emission at 268 nm. This work offers a practical approach for optimizing the electrical performance of AlGaN-based DUV LEDs and provides insights into the co-optimization of p- and n-type contacts in device fabrication. 

\begin{acknowledgments}
This work was supported by the Army Research Office
under Grant No. W911NF2220177 (characterization); ULTRA, an
Energy Frontier Research Center funded by the U.S. Department of Energy (DOE); SUPREME (modeling), one of seven centers in JUMP 2.0, a Semiconductor Research Corporation (SRC) program sponsored by DARPA; and the DARPA UWBGS program. This work was performed in part at the Cornell NanoScale Facility, an NNCI member supported by NSF Grant NNCI 2025233. 
\end{acknowledgments}

\section*{author declarations}
H.W. Huang and S. Agrawal contributed equally to this work.
\subsection*{Conflict of Interest}
The authors have no conflict of interest to declare.

\section*{Data Availability Statement}
The data that support the findings of this study are available from the corresponding authors upon reasonable request.

\nocite{*}
\bibliography{contacts_study.bib}

\begin{thebibliography}{41}%
\makeatletter
\providecommand \@ifxundefined [1]{%
 \@ifx{#1\undefined}
}%
\providecommand \@ifnum [1]{%
 \ifnum #1\expandafter \@firstoftwo
 \else \expandafter \@secondoftwo
 \fi
}%
\providecommand \@ifx [1]{%
 \ifx #1\expandafter \@firstoftwo
 \else \expandafter \@secondoftwo
 \fi
}%
\providecommand \natexlab [1]{#1}%
\providecommand \enquote  [1]{``#1''}%
\providecommand \bibnamefont  [1]{#1}%
\providecommand \bibfnamefont [1]{#1}%
\providecommand \citenamefont [1]{#1}%
\providecommand \href@noop [0]{\@secondoftwo}%
\providecommand \href [0]{\begingroup \@sanitize@url \@href}%
\providecommand \@href[1]{\@@startlink{#1}\@@href}%
\providecommand \@@href[1]{\endgroup#1\@@endlink}%
\providecommand \@sanitize@url [0]{\catcode `\\12\catcode `\$12\catcode `\&12\catcode `\#12\catcode `\^12\catcode `\_12\catcode `\%12\relax}%
\providecommand \@@startlink[1]{}%
\providecommand \@@endlink[0]{}%
\providecommand \url  [0]{\begingroup\@sanitize@url \@url }%
\providecommand \@url [1]{\endgroup\@href {#1}{\urlprefix }}%
\providecommand \urlprefix  [0]{URL }%
\providecommand \Eprint [0]{\href }%
\providecommand \doibase [0]{https://doi.org/}%
\providecommand \selectlanguage [0]{\@gobble}%
\providecommand \bibinfo  [0]{\@secondoftwo}%
\providecommand \bibfield  [0]{\@secondoftwo}%
\providecommand \translation [1]{[#1]}%
\providecommand \BibitemOpen [0]{}%
\providecommand \bibitemStop [0]{}%
\providecommand \bibitemNoStop [0]{.\EOS\space}%
\providecommand \EOS [0]{\spacefactor3000\relax}%
\providecommand \BibitemShut  [1]{\csname bibitem#1\endcsname}%
\let\auto@bib@innerbib\@empty
\bibitem [{\citenamefont {Zhang}\ \emph {et~al.}(2019)\citenamefont {Zhang}, \citenamefont {Kushimoto}, \citenamefont {Sakai}, \citenamefont {Sugiyama}, \citenamefont {Schowalter}, \citenamefont {Sasaoka},\ and\ \citenamefont {Amano}}]{zhang2019271}%
  \BibitemOpen
  \bibfield  {author} {\bibinfo {author} {\bibfnamefont {Z.}~\bibnamefont {Zhang}}, \bibinfo {author} {\bibfnamefont {M.}~\bibnamefont {Kushimoto}}, \bibinfo {author} {\bibfnamefont {T.}~\bibnamefont {Sakai}}, \bibinfo {author} {\bibfnamefont {N.}~\bibnamefont {Sugiyama}}, \bibinfo {author} {\bibfnamefont {L.~J.}\ \bibnamefont {Schowalter}}, \bibinfo {author} {\bibfnamefont {C.}~\bibnamefont {Sasaoka}},\ and\ \bibinfo {author} {\bibfnamefont {H.}~\bibnamefont {Amano}},\ }\bibfield  {title} {\enquote {\bibinfo {title} {A 271.8 nm deep-ultraviolet laser diode for room temperature operation},}\ }\href {https://doi.org/10.7567/1882-0786/ab50e0} {\bibfield  {journal} {\bibinfo  {journal} {Applied Physics Express}\ }\textbf {\bibinfo {volume} {12}},\ \bibinfo {pages} {124003} (\bibinfo {year} {2019})}\BibitemShut {NoStop}%
\bibitem [{\citenamefont {Lang}\ \emph {et~al.}(2025)\citenamefont {Lang}, \citenamefont {Xu}, \citenamefont {Wang}, \citenamefont {Zhang}, \citenamefont {Fang}, \citenamefont {Zhang}, \citenamefont {Guo}, \citenamefont {Ji}, \citenamefont {Ji}, \citenamefont {Tan}, \citenamefont {Wu}, \citenamefont {Yang}, \citenamefont {Kang}, \citenamefont {Qin}, \citenamefont {Tang}, \citenamefont {Wang}, \citenamefont {Ge},\ and\ \citenamefont {Shen}}]{lang2025progress}%
  \BibitemOpen
  \bibfield  {author} {\bibinfo {author} {\bibfnamefont {J.}~\bibnamefont {Lang}}, \bibinfo {author} {\bibfnamefont {F.}~\bibnamefont {Xu}}, \bibinfo {author} {\bibfnamefont {J.}~\bibnamefont {Wang}}, \bibinfo {author} {\bibfnamefont {L.}~\bibnamefont {Zhang}}, \bibinfo {author} {\bibfnamefont {X.}~\bibnamefont {Fang}}, \bibinfo {author} {\bibfnamefont {Z.}~\bibnamefont {Zhang}}, \bibinfo {author} {\bibfnamefont {X.}~\bibnamefont {Guo}}, \bibinfo {author} {\bibfnamefont {C.}~\bibnamefont {Ji}}, \bibinfo {author} {\bibfnamefont {C.}~\bibnamefont {Ji}}, \bibinfo {author} {\bibfnamefont {F.}~\bibnamefont {Tan}}, \bibinfo {author} {\bibfnamefont {Y.}~\bibnamefont {Wu}}, \bibinfo {author} {\bibfnamefont {X.}~\bibnamefont {Yang}}, \bibinfo {author} {\bibfnamefont {X.}~\bibnamefont {Kang}}, \bibinfo {author} {\bibfnamefont {Z.}~\bibnamefont {Qin}}, \bibinfo {author} {\bibfnamefont {N.}~\bibnamefont {Tang}}, \bibinfo {author} {\bibfnamefont {X.}~\bibnamefont {Wang}}, \bibinfo {author} {\bibfnamefont {W.}~\bibnamefont
  {Ge}},\ and\ \bibinfo {author} {\bibfnamefont {B.}~\bibnamefont {Shen}},\ }\bibfield  {title} {\enquote {\bibinfo {title} {Progress in performance of {AlGaN}-based ultraviolet light emitting diodes},}\ }\href {https://doi.org/10.1002/aelm.202300840} {\bibfield  {journal} {\bibinfo  {journal} {Advanced Electronic Materials}\ }\textbf {\bibinfo {volume} {11}},\ \bibinfo {pages} {2300840} (\bibinfo {year} {2025})}\BibitemShut {NoStop}%
\bibitem [{\citenamefont {Kneissl}\ and\ \citenamefont {Rass}(2016)}]{kneissl2016iii}%
  \BibitemOpen
  \bibfield  {author} {\bibinfo {author} {\bibfnamefont {M.}~\bibnamefont {Kneissl}}\ and\ \bibinfo {author} {\bibfnamefont {J.}~\bibnamefont {Rass}},\ }\href@noop {} {\emph {\bibinfo {title} {III-Nitride Ultraviolet Emitters}}}\ (\bibinfo  {publisher} {Springer},\ \bibinfo {year} {2016})\BibitemShut {NoStop}%
\bibitem [{\citenamefont {Sulmoni}\ \emph {et~al.}(2020)\citenamefont {Sulmoni}, \citenamefont {Mehnke}, \citenamefont {Mogilatenko}, \citenamefont {Guttmann}, \citenamefont {Wernicke},\ and\ \citenamefont {Kneissl}}]{sulmoni2020electrical}%
  \BibitemOpen
  \bibfield  {author} {\bibinfo {author} {\bibfnamefont {L.}~\bibnamefont {Sulmoni}}, \bibinfo {author} {\bibfnamefont {F.}~\bibnamefont {Mehnke}}, \bibinfo {author} {\bibfnamefont {A.}~\bibnamefont {Mogilatenko}}, \bibinfo {author} {\bibfnamefont {M.}~\bibnamefont {Guttmann}}, \bibinfo {author} {\bibfnamefont {T.}~\bibnamefont {Wernicke}},\ and\ \bibinfo {author} {\bibfnamefont {M.}~\bibnamefont {Kneissl}},\ }\bibfield  {title} {\enquote {\bibinfo {title} {Electrical properties and microstructure formation of {V/Al}-based n-contacts on high {Al} mole fraction n-{AlGaN} layers},}\ }\href {https://doi.org/10.1364/PRJ.391075} {\bibfield  {journal} {\bibinfo  {journal} {Photonics Research}\ }\textbf {\bibinfo {volume} {8}},\ \bibinfo {pages} {1381--1387} (\bibinfo {year} {2020})}\BibitemShut {NoStop}%
\bibitem [{\citenamefont {Taniyasu}, \citenamefont {Kasu},\ and\ \citenamefont {Kobayashi}(2002)}]{taniyasu2002intentional}%
  \BibitemOpen
  \bibfield  {author} {\bibinfo {author} {\bibfnamefont {Y.}~\bibnamefont {Taniyasu}}, \bibinfo {author} {\bibfnamefont {M.}~\bibnamefont {Kasu}},\ and\ \bibinfo {author} {\bibfnamefont {N.}~\bibnamefont {Kobayashi}},\ }\bibfield  {title} {\enquote {\bibinfo {title} {Intentional control of n-type conduction for {Si-doped AlN and Al$_{x}$Ga$_{1-x}$N (0.42< x< 1)}},}\ }\href {https://doi.org/10.1063/1.1499738} {\bibfield  {journal} {\bibinfo  {journal} {Applied Physics Letters}\ }\textbf {\bibinfo {volume} {81}},\ \bibinfo {pages} {1255--1257} (\bibinfo {year} {2002})}\BibitemShut {NoStop}%
\bibitem [{\citenamefont {Nagata}\ \emph {et~al.}(2017)\citenamefont {Nagata}, \citenamefont {Senga}, \citenamefont {Iwaya}, \citenamefont {Takeuchi}, \citenamefont {Kamiyama},\ and\ \citenamefont {Akasaki}}]{nagata2017reduction}%
  \BibitemOpen
  \bibfield  {author} {\bibinfo {author} {\bibfnamefont {N.}~\bibnamefont {Nagata}}, \bibinfo {author} {\bibfnamefont {T.}~\bibnamefont {Senga}}, \bibinfo {author} {\bibfnamefont {M.}~\bibnamefont {Iwaya}}, \bibinfo {author} {\bibfnamefont {T.}~\bibnamefont {Takeuchi}}, \bibinfo {author} {\bibfnamefont {S.}~\bibnamefont {Kamiyama}},\ and\ \bibinfo {author} {\bibfnamefont {I.}~\bibnamefont {Akasaki}},\ }\bibfield  {title} {\enquote {\bibinfo {title} {Reduction of contact resistance in {V-based electrode for high AlN molar fraction n-type AlGaN by using thin SiNx} intermediate layer},}\ }\href {https://doi.org/10.1002/pssc.201600243} {\bibfield  {journal} {\bibinfo  {journal} {Physica Status Solidi C}\ }\textbf {\bibinfo {volume} {14}},\ \bibinfo {pages} {1600243} (\bibinfo {year} {2017})}\BibitemShut {NoStop}%
\bibitem [{\citenamefont {Bai}\ \emph {et~al.}(2012)\citenamefont {Bai}, \citenamefont {Liu}, \citenamefont {Shen}, \citenamefont {Ma}, \citenamefont {Liu},\ and\ \citenamefont {Guo}}]{bai2012effect}%
  \BibitemOpen
  \bibfield  {author} {\bibinfo {author} {\bibfnamefont {Y.}~\bibnamefont {Bai}}, \bibinfo {author} {\bibfnamefont {J.}~\bibnamefont {Liu}}, \bibinfo {author} {\bibfnamefont {H.}~\bibnamefont {Shen}}, \bibinfo {author} {\bibfnamefont {P.}~\bibnamefont {Ma}}, \bibinfo {author} {\bibfnamefont {X.}~\bibnamefont {Liu}},\ and\ \bibinfo {author} {\bibfnamefont {L.}~\bibnamefont {Guo}},\ }\bibfield  {title} {\enquote {\bibinfo {title} {Effect of annealing on the characteristics of {Pd/Au} contacts to p-type {GaN/Al$_{0.45}$Ga$_{0.55}$N}},}\ }\href {https://link.springer.com/article/10.1007/s11664-012-2183-6} {\bibfield  {journal} {\bibinfo  {journal} {Journal of Electronic Materials}\ }\textbf {\bibinfo {volume} {41}},\ \bibinfo {pages} {3021--3026} (\bibinfo {year} {2012})}\BibitemShut {NoStop}%
\bibitem [{\citenamefont {Greco}, \citenamefont {Iucolano},\ and\ \citenamefont {Roccaforte}(2016)}]{greco2016ohmic}%
  \BibitemOpen
  \bibfield  {author} {\bibinfo {author} {\bibfnamefont {G.}~\bibnamefont {Greco}}, \bibinfo {author} {\bibfnamefont {F.}~\bibnamefont {Iucolano}},\ and\ \bibinfo {author} {\bibfnamefont {F.}~\bibnamefont {Roccaforte}},\ }\bibfield  {title} {\enquote {\bibinfo {title} {Ohmic contacts to gallium nitride materials},}\ }\href {https://doi.org/10.1016/j.apsusc.2016.04.016} {\bibfield  {journal} {\bibinfo  {journal} {Applied Surface Science}\ }\textbf {\bibinfo {volume} {383}},\ \bibinfo {pages} {324--345} (\bibinfo {year} {2016})}\BibitemShut {NoStop}%
\bibitem [{\citenamefont {Agrawal}\ \emph {et~al.}(2024)\citenamefont {Agrawal}, \citenamefont {van Deurzen}, \citenamefont {Encomendero}, \citenamefont {Dill}, \citenamefont {Wei Sheena~Huang}, \citenamefont {Protasenko}, \citenamefont {Xing},\ and\ \citenamefont {Jena}}]{agrawal2024ultrawide}%
  \BibitemOpen
  \bibfield  {author} {\bibinfo {author} {\bibfnamefont {S.}~\bibnamefont {Agrawal}}, \bibinfo {author} {\bibfnamefont {L.}~\bibnamefont {van Deurzen}}, \bibinfo {author} {\bibfnamefont {J.}~\bibnamefont {Encomendero}}, \bibinfo {author} {\bibfnamefont {J.~E.}\ \bibnamefont {Dill}}, \bibinfo {author} {\bibfnamefont {H.}~\bibnamefont {Wei Sheena~Huang}}, \bibinfo {author} {\bibfnamefont {V.}~\bibnamefont {Protasenko}}, \bibinfo {author} {\bibfnamefont {H.~G.}\ \bibnamefont {Xing}},\ and\ \bibinfo {author} {\bibfnamefont {D.}~\bibnamefont {Jena}},\ }\bibfield  {title} {\enquote {\bibinfo {title} {Ultrawide bandgap semiconductor heterojunction p--n diodes with distributed polarization-doped p-type {AlGaN} layers on bulk {AlN} substrates},}\ }\href {https://doi.org/10.1063/5.0189419} {\bibfield  {journal} {\bibinfo  {journal} {Applied Physics Letters}\ }\textbf {\bibinfo {volume} {124}},\ \bibinfo {pages} {102109} (\bibinfo {year} {2024})}\BibitemShut {NoStop}%
\bibitem [{\citenamefont {Hao}\ \emph {et~al.}(2017)\citenamefont {Hao}, \citenamefont {Taniguchi}, \citenamefont {Tamari},\ and\ \citenamefont {Inoue}}]{hao2017improved}%
  \BibitemOpen
  \bibfield  {author} {\bibinfo {author} {\bibfnamefont {G.-D.}\ \bibnamefont {Hao}}, \bibinfo {author} {\bibfnamefont {M.}~\bibnamefont {Taniguchi}}, \bibinfo {author} {\bibfnamefont {N.}~\bibnamefont {Tamari}},\ and\ \bibinfo {author} {\bibfnamefont {S.-I.}\ \bibnamefont {Inoue}},\ }\bibfield  {title} {\enquote {\bibinfo {title} {Improved turn-on and operating voltages in {AlGaN}-based deep-ultraviolet light-emitting diodes},}\ }\href {https://doi.org/10.1007/s11664-017-5622-6} {\bibfield  {journal} {\bibinfo  {journal} {Journal of Electronic Materials}\ }\textbf {\bibinfo {volume} {46}},\ \bibinfo {pages} {5677--5683} (\bibinfo {year} {2017})}\BibitemShut {NoStop}%
\bibitem [{\citenamefont {Zhang}\ \emph {et~al.}(2022)\citenamefont {Zhang}, \citenamefont {Kushimoto}, \citenamefont {Yoshikawa}, \citenamefont {Aoto}, \citenamefont {Sasaoka}, \citenamefont {Schowalter},\ and\ \citenamefont {Amano}}]{zhang2022key}%
  \BibitemOpen
  \bibfield  {author} {\bibinfo {author} {\bibfnamefont {Z.}~\bibnamefont {Zhang}}, \bibinfo {author} {\bibfnamefont {M.}~\bibnamefont {Kushimoto}}, \bibinfo {author} {\bibfnamefont {A.}~\bibnamefont {Yoshikawa}}, \bibinfo {author} {\bibfnamefont {K.}~\bibnamefont {Aoto}}, \bibinfo {author} {\bibfnamefont {C.}~\bibnamefont {Sasaoka}}, \bibinfo {author} {\bibfnamefont {L.~J.}\ \bibnamefont {Schowalter}},\ and\ \bibinfo {author} {\bibfnamefont {H.}~\bibnamefont {Amano}},\ }\bibfield  {title} {\enquote {\bibinfo {title} {Key temperature-dependent characteristics of {AlGaN}-based {UV-C} laser diode and demonstration of room-temperature continuous-wave lasing},}\ }\href {https://doi.org/10.1063/5.0124480} {\bibfield  {journal} {\bibinfo  {journal} {Applied Physics Letters}\ }\textbf {\bibinfo {volume} {121}},\ \bibinfo {pages} {222103} (\bibinfo {year} {2022})}\BibitemShut {NoStop}%
\bibitem [{\citenamefont {Dill}\ \emph {et~al.}(2025)\citenamefont {Dill}, \citenamefont {Shoemaker}, \citenamefont {Nomoto}, \citenamefont {Encomendero}, \citenamefont {Zhang}, \citenamefont {Chang}, \citenamefont {Chen}, \citenamefont {Giustino}, \citenamefont {Goodnick}, \citenamefont {Jena},\ and\ \citenamefont {Xing}}]{dill2025velocity}%
  \BibitemOpen
  \bibfield  {author} {\bibinfo {author} {\bibfnamefont {J.~E.}\ \bibnamefont {Dill}}, \bibinfo {author} {\bibfnamefont {J.}~\bibnamefont {Shoemaker}}, \bibinfo {author} {\bibfnamefont {K.}~\bibnamefont {Nomoto}}, \bibinfo {author} {\bibfnamefont {J.}~\bibnamefont {Encomendero}}, \bibinfo {author} {\bibfnamefont {Z.}~\bibnamefont {Zhang}}, \bibinfo {author} {\bibfnamefont {C.~F.}\ \bibnamefont {Chang}}, \bibinfo {author} {\bibfnamefont {J.-C.}\ \bibnamefont {Chen}}, \bibinfo {author} {\bibfnamefont {F.}~\bibnamefont {Giustino}}, \bibinfo {author} {\bibfnamefont {S.}~\bibnamefont {Goodnick}}, \bibinfo {author} {\bibfnamefont {D.}~\bibnamefont {Jena}},\ and\ \bibinfo {author} {\bibfnamefont {H.~G.}\ \bibnamefont {Xing}},\ }\bibfield  {title} {\enquote {\bibinfo {title} {Velocity-field measurements in a {GaN/AlN} two-dimensional hole gas},}\ }\href {https://doi.org/10.1063/5.0276423} {\bibfield  {journal} {\bibinfo  {journal} {Applied Physics Letters}\ }\textbf {\bibinfo {volume} {127}},\ \bibinfo {pages}
  {032105} (\bibinfo {year} {2025})}\BibitemShut {NoStop}%
\bibitem [{\citenamefont {Bader}\ \emph {et~al.}(2019)\citenamefont {Bader}, \citenamefont {Chaudhuri}, \citenamefont {Hickman}, \citenamefont {Nomoto}, \citenamefont {Bharadwaj}, \citenamefont {Then}, \citenamefont {Xing},\ and\ \citenamefont {Jena}}]{bader2019gan}%
  \BibitemOpen
  \bibfield  {author} {\bibinfo {author} {\bibfnamefont {S.}~\bibnamefont {Bader}}, \bibinfo {author} {\bibfnamefont {R.}~\bibnamefont {Chaudhuri}}, \bibinfo {author} {\bibfnamefont {A.}~\bibnamefont {Hickman}}, \bibinfo {author} {\bibfnamefont {K.}~\bibnamefont {Nomoto}}, \bibinfo {author} {\bibfnamefont {S.}~\bibnamefont {Bharadwaj}}, \bibinfo {author} {\bibfnamefont {H.}~\bibnamefont {Then}}, \bibinfo {author} {\bibfnamefont {H.}~\bibnamefont {Xing}},\ and\ \bibinfo {author} {\bibfnamefont {D.}~\bibnamefont {Jena}},\ }\bibfield  {title} {\enquote {\bibinfo {title} {{GaN/AlN} schottky-gate p-channel {HFETs with InGaN} contacts and 100 ma/mm on-current},}\ }in\ \href {https://doi.org/10.1109/IEDM19573.2019.8993532} {\emph {\bibinfo {booktitle} {2019 IEEE International Electron Devices Meeting (IEDM)}}}\ (\bibinfo {organization} {IEEE},\ \bibinfo {year} {2019})\ pp.\ \bibinfo {pages} {4--5}\BibitemShut {NoStop}%
\bibitem [{\citenamefont {Kumakura}, \citenamefont {Makimoto},\ and\ \citenamefont {Kobayashi}(2003)}]{kumakura2003ohmic}%
  \BibitemOpen
  \bibfield  {author} {\bibinfo {author} {\bibfnamefont {K.}~\bibnamefont {Kumakura}}, \bibinfo {author} {\bibfnamefont {T.}~\bibnamefont {Makimoto}},\ and\ \bibinfo {author} {\bibfnamefont {N.}~\bibnamefont {Kobayashi}},\ }\bibfield  {title} {\enquote {\bibinfo {title} {Ohmic contact to {p-GaN using a strained InGaN} contact layer and its thermal stability},}\ }\href {https://doi.org/10.1143/JJAP.42.2254} {\bibfield  {journal} {\bibinfo  {journal} {Japanese Journal of Applied Physics}\ }\textbf {\bibinfo {volume} {42}},\ \bibinfo {pages} {2254} (\bibinfo {year} {2003})}\BibitemShut {NoStop}%
\bibitem [{\citenamefont {Snider}(2024)}]{snider1dpoisson}%
  \BibitemOpen
  \bibfield  {author} {\bibinfo {author} {\bibfnamefont {G.}~\bibnamefont {Snider}},\ }\href@noop {} {\enquote {\bibinfo {title} {1{D Poisson/Schrödinger Solver}},}\ }\bibinfo {howpublished} {Available at: https://www3.nd.edu/~gsnider/} (\bibinfo {year} {2024}),\ \bibinfo {note} {version Beta 8, University of Notre Dame}\BibitemShut {NoStop}%
\bibitem [{\citenamefont {Alam}\ \emph {et~al.}(2020)\citenamefont {Alam}, \citenamefont {Zubialevich}, \citenamefont {Ghafary},\ and\ \citenamefont {Parbrook}}]{alam2020bandgap}%
  \BibitemOpen
  \bibfield  {author} {\bibinfo {author} {\bibfnamefont {S.~N.}\ \bibnamefont {Alam}}, \bibinfo {author} {\bibfnamefont {V.~Z.}\ \bibnamefont {Zubialevich}}, \bibinfo {author} {\bibfnamefont {B.}~\bibnamefont {Ghafary}},\ and\ \bibinfo {author} {\bibfnamefont {P.~J.}\ \bibnamefont {Parbrook}},\ }\bibfield  {title} {\enquote {\bibinfo {title} {Bandgap and refractive index estimates of {InAlN} and related nitrides across their full composition ranges},}\ }\href {https://www.nature.com/articles/s41598-020-73160-7} {\bibfield  {journal} {\bibinfo  {journal} {Scientific Reports}\ }\textbf {\bibinfo {volume} {10}},\ \bibinfo {pages} {16205} (\bibinfo {year} {2020})}\BibitemShut {NoStop}%
\bibitem [{\citenamefont {Lin}\ \emph {et~al.}(2012)\citenamefont {Lin}, \citenamefont {Kuo}, \citenamefont {Liu}, \citenamefont {Liang}, \citenamefont {Cheng}, \citenamefont {Lin}, \citenamefont {Tang}, \citenamefont {Chang}, \citenamefont {Chen},\ and\ \citenamefont {Gwo}}]{lin2012experimental}%
  \BibitemOpen
  \bibfield  {author} {\bibinfo {author} {\bibfnamefont {S.-C.}\ \bibnamefont {Lin}}, \bibinfo {author} {\bibfnamefont {C.-T.}\ \bibnamefont {Kuo}}, \bibinfo {author} {\bibfnamefont {X.}~\bibnamefont {Liu}}, \bibinfo {author} {\bibfnamefont {L.-Y.}\ \bibnamefont {Liang}}, \bibinfo {author} {\bibfnamefont {C.-H.}\ \bibnamefont {Cheng}}, \bibinfo {author} {\bibfnamefont {C.-H.}\ \bibnamefont {Lin}}, \bibinfo {author} {\bibfnamefont {S.-J.}\ \bibnamefont {Tang}}, \bibinfo {author} {\bibfnamefont {L.-Y.}\ \bibnamefont {Chang}}, \bibinfo {author} {\bibfnamefont {C.-H.}\ \bibnamefont {Chen}},\ and\ \bibinfo {author} {\bibfnamefont {S.}~\bibnamefont {Gwo}},\ }\bibfield  {title} {\enquote {\bibinfo {title} {Experimental determination of electron affinities for {InN and GaN} polar surfaces},}\ }\href {https://doi.org/10.1143/APEX.5.031003} {\bibfield  {journal} {\bibinfo  {journal} {Applied Physics Express}\ }\textbf {\bibinfo {volume} {5}},\ \bibinfo {pages} {031003} (\bibinfo {year} {2012})}\BibitemShut {NoStop}%
\bibitem [{\citenamefont {Zhao}\ \emph {et~al.}(2017)\citenamefont {Zhao}, \citenamefont {Wei}, \citenamefont {Chen}, \citenamefont {Wang},\ and\ \citenamefont {Wang}}]{zhao2017activation}%
  \BibitemOpen
  \bibfield  {author} {\bibinfo {author} {\bibfnamefont {C.-Z.}\ \bibnamefont {Zhao}}, \bibinfo {author} {\bibfnamefont {T.}~\bibnamefont {Wei}}, \bibinfo {author} {\bibfnamefont {L.-Y.}\ \bibnamefont {Chen}}, \bibinfo {author} {\bibfnamefont {S.-S.}\ \bibnamefont {Wang}},\ and\ \bibinfo {author} {\bibfnamefont {J.}~\bibnamefont {Wang}},\ }\bibfield  {title} {\enquote {\bibinfo {title} {The activation energy for {Mg} acceptor in the {Ga-rich InGaN} alloys},}\ }\href {https://doi.org/10.1016/j.spmi.2016.12.024} {\bibfield  {journal} {\bibinfo  {journal} {Superlattices and Microstructures}\ }\textbf {\bibinfo {volume} {102}},\ \bibinfo {pages} {40--44} (\bibinfo {year} {2017})}\BibitemShut {NoStop}%
\bibitem [{\citenamefont {Sze}, \citenamefont {Li},\ and\ \citenamefont {Ng}(2021)}]{sze2021physics}%
  \BibitemOpen
  \bibfield  {author} {\bibinfo {author} {\bibfnamefont {S.~M.}\ \bibnamefont {Sze}}, \bibinfo {author} {\bibfnamefont {Y.}~\bibnamefont {Li}},\ and\ \bibinfo {author} {\bibfnamefont {K.~K.}\ \bibnamefont {Ng}},\ }\href@noop {} {\emph {\bibinfo {title} {Physics of Semiconductor Devices}}}\ (\bibinfo  {publisher} {John Wiley \& Sons},\ \bibinfo {year} {2021})\BibitemShut {NoStop}%
\bibitem [{\citenamefont {Dai}\ \emph {et~al.}(2021)\citenamefont {Dai}, \citenamefont {Mai}, \citenamefont {Wu}, \citenamefont {Peng}, \citenamefont {Liu}, \citenamefont {Wen}, \citenamefont {Chou}, \citenamefont {Ho},\ and\ \citenamefont {Wang}}]{dai2021high}%
  \BibitemOpen
  \bibfield  {author} {\bibinfo {author} {\bibfnamefont {J.-J.}\ \bibnamefont {Dai}}, \bibinfo {author} {\bibfnamefont {T.~T.}\ \bibnamefont {Mai}}, \bibinfo {author} {\bibfnamefont {S.-K.}\ \bibnamefont {Wu}}, \bibinfo {author} {\bibfnamefont {J.-R.}\ \bibnamefont {Peng}}, \bibinfo {author} {\bibfnamefont {C.-W.}\ \bibnamefont {Liu}}, \bibinfo {author} {\bibfnamefont {H.-C.}\ \bibnamefont {Wen}}, \bibinfo {author} {\bibfnamefont {W.-C.}\ \bibnamefont {Chou}}, \bibinfo {author} {\bibfnamefont {H.-C.}\ \bibnamefont {Ho}},\ and\ \bibinfo {author} {\bibfnamefont {W.-F.}\ \bibnamefont {Wang}},\ }\bibfield  {title} {\enquote {\bibinfo {title} {High hole concentration and diffusion suppression of heavily {Mg-doped p-GaN for application in enhanced-mode GaN HEMT}},}\ }\href {https://doi.org/10.3390/nano11071766} {\bibfield  {journal} {\bibinfo  {journal} {Nanomaterials}\ }\textbf {\bibinfo {volume} {11}},\ \bibinfo {pages} {1766} (\bibinfo {year} {2021})}\BibitemShut {NoStop}%
\bibitem [{\citenamefont {Chaudhuri}\ \emph {et~al.}(2019)\citenamefont {Chaudhuri}, \citenamefont {Bader}, \citenamefont {Chen}, \citenamefont {Muller}, \citenamefont {Xing},\ and\ \citenamefont {Jena}}]{chaudhuri2019polarization}%
  \BibitemOpen
  \bibfield  {author} {\bibinfo {author} {\bibfnamefont {R.}~\bibnamefont {Chaudhuri}}, \bibinfo {author} {\bibfnamefont {S.~J.}\ \bibnamefont {Bader}}, \bibinfo {author} {\bibfnamefont {Z.}~\bibnamefont {Chen}}, \bibinfo {author} {\bibfnamefont {D.~A.}\ \bibnamefont {Muller}}, \bibinfo {author} {\bibfnamefont {H.~G.}\ \bibnamefont {Xing}},\ and\ \bibinfo {author} {\bibfnamefont {D.}~\bibnamefont {Jena}},\ }\bibfield  {title} {\enquote {\bibinfo {title} {A polarization-induced {2D} hole gas in undoped gallium nitride quantum wells},}\ }\href {https://doi.org/DOI: 10.1126/science.aau8623} {\bibfield  {journal} {\bibinfo  {journal} {Science}\ }\textbf {\bibinfo {volume} {365}},\ \bibinfo {pages} {1454--1457} (\bibinfo {year} {2019})}\BibitemShut {NoStop}%
\bibitem [{\citenamefont {Chaudhuri}\ \emph {et~al.}(2022)\citenamefont {Chaudhuri}, \citenamefont {Zhang}, \citenamefont {Xing},\ and\ \citenamefont {Jena}}]{chaudhuri2022very}%
  \BibitemOpen
  \bibfield  {author} {\bibinfo {author} {\bibfnamefont {R.}~\bibnamefont {Chaudhuri}}, \bibinfo {author} {\bibfnamefont {Z.}~\bibnamefont {Zhang}}, \bibinfo {author} {\bibfnamefont {H.~G.}\ \bibnamefont {Xing}},\ and\ \bibinfo {author} {\bibfnamefont {D.}~\bibnamefont {Jena}},\ }\bibfield  {title} {\enquote {\bibinfo {title} {Very high density (> 10$^{14}$ cm$^{-2}$) polarization-induced {2D} hole gases observed in undoped pseudomorphic {InGaN/AlN} heterostructures},}\ }\href {https://doi.org/10.1002/aelm.202101120} {\bibfield  {journal} {\bibinfo  {journal} {Advanced Electronic Materials}\ }\textbf {\bibinfo {volume} {8}},\ \bibinfo {pages} {2101120} (\bibinfo {year} {2022})}\BibitemShut {NoStop}%
\bibitem [{\citenamefont {Kumakura}, \citenamefont {Makimoto},\ and\ \citenamefont {Kobayashi}(2000)}]{kumakura2000activation}%
  \BibitemOpen
  \bibfield  {author} {\bibinfo {author} {\bibfnamefont {K.}~\bibnamefont {Kumakura}}, \bibinfo {author} {\bibfnamefont {T.}~\bibnamefont {Makimoto}},\ and\ \bibinfo {author} {\bibfnamefont {N.}~\bibnamefont {Kobayashi}},\ }\bibfield  {title} {\enquote {\bibinfo {title} {Activation energy and electrical activity of {Mg in Mg-doped} {In$_{x}$Ga$_{1-x}$N} (x< 0.2)},}\ }\href {https://iopscience.iop.org/article/10.1143/JJAP.39.L337} {\bibfield  {journal} {\bibinfo  {journal} {Japanese Journal of Applied Physics}\ }\textbf {\bibinfo {volume} {39}},\ \bibinfo {pages} {L337} (\bibinfo {year} {2000})}\BibitemShut {NoStop}%
\bibitem [{\citenamefont {Lee}(2020)}]{KevinLee_thesis}%
  \BibitemOpen
  \bibfield  {author} {\bibinfo {author} {\bibfnamefont {K.}~\bibnamefont {Lee}},\ }\emph {\bibinfo {title} {Improving Efficiency of Visible and Deep UV LEDs and Lasers}},\ \href@noop {} {\bibinfo {type} {Doctoral dissertation}},\ \bibinfo  {school} {Cornell University}, \bibinfo {address} {Ithaca, New York 14853, USA} (\bibinfo {year} {2020})\BibitemShut {NoStop}%
\bibitem [{\citenamefont {Kumar}\ \emph {et~al.}(2023)\citenamefont {Kumar}, \citenamefont {Berg}, \citenamefont {Wang}, \citenamefont {Salter},\ and\ \citenamefont {Ramvall}}]{kumar2023growth}%
  \BibitemOpen
  \bibfield  {author} {\bibinfo {author} {\bibfnamefont {A.}~\bibnamefont {Kumar}}, \bibinfo {author} {\bibfnamefont {M.}~\bibnamefont {Berg}}, \bibinfo {author} {\bibfnamefont {Q.}~\bibnamefont {Wang}}, \bibinfo {author} {\bibfnamefont {M.}~\bibnamefont {Salter}},\ and\ \bibinfo {author} {\bibfnamefont {P.}~\bibnamefont {Ramvall}},\ }\bibfield  {title} {\enquote {\bibinfo {title} {Growth of p-type {GaN}--the role of oxygen in activation of {Mg}-doping},}\ }\href {https://doi.org/10.1016/j.pedc.2023.100036} {\bibfield  {journal} {\bibinfo  {journal} {Power Electronic Devices and Components}\ }\textbf {\bibinfo {volume} {5}},\ \bibinfo {pages} {100036} (\bibinfo {year} {2023})}\BibitemShut {NoStop}%
\bibitem [{\citenamefont {Li}\ \emph {et~al.}(2017)\citenamefont {Li}, \citenamefont {Nomoto}, \citenamefont {Pilla}, \citenamefont {Pan}, \citenamefont {Gao}, \citenamefont {Jena},\ and\ \citenamefont {Xing}}]{li2017design}%
  \BibitemOpen
  \bibfield  {author} {\bibinfo {author} {\bibfnamefont {W.}~\bibnamefont {Li}}, \bibinfo {author} {\bibfnamefont {K.}~\bibnamefont {Nomoto}}, \bibinfo {author} {\bibfnamefont {M.}~\bibnamefont {Pilla}}, \bibinfo {author} {\bibfnamefont {M.}~\bibnamefont {Pan}}, \bibinfo {author} {\bibfnamefont {X.}~\bibnamefont {Gao}}, \bibinfo {author} {\bibfnamefont {D.}~\bibnamefont {Jena}},\ and\ \bibinfo {author} {\bibfnamefont {H.~G.}\ \bibnamefont {Xing}},\ }\bibfield  {title} {\enquote {\bibinfo {title} {Design and realization of {GaN} trench junction-barrier-{Schottky}-diodes},}\ }\href {https://doi.org/10.1109/TED.2017.2662702} {\bibfield  {journal} {\bibinfo  {journal} {IEEE Transactions on Electron Devices}\ }\textbf {\bibinfo {volume} {64}},\ \bibinfo {pages} {1635--1641} (\bibinfo {year} {2017})}\BibitemShut {NoStop}%
\bibitem [{\citenamefont {Ho}\ \emph {et~al.}(1999)\citenamefont {Ho}, \citenamefont {Jong}, \citenamefont {Chiu}, \citenamefont {Huang}, \citenamefont {Chen},\ and\ \citenamefont {Shih}}]{ho1999low}%
  \BibitemOpen
  \bibfield  {author} {\bibinfo {author} {\bibfnamefont {J.-K.}\ \bibnamefont {Ho}}, \bibinfo {author} {\bibfnamefont {C.-S.}\ \bibnamefont {Jong}}, \bibinfo {author} {\bibfnamefont {C.~C.}\ \bibnamefont {Chiu}}, \bibinfo {author} {\bibfnamefont {C.-N.}\ \bibnamefont {Huang}}, \bibinfo {author} {\bibfnamefont {C.-Y.}\ \bibnamefont {Chen}},\ and\ \bibinfo {author} {\bibfnamefont {K.-K.}\ \bibnamefont {Shih}},\ }\bibfield  {title} {\enquote {\bibinfo {title} {Low-resistance ohmic contacts to p-type {GaN}},}\ }\href {https://doi.org/10.1063/1.123546} {\bibfield  {journal} {\bibinfo  {journal} {Applied Physics Letters}\ }\textbf {\bibinfo {volume} {74}},\ \bibinfo {pages} {1275--1277} (\bibinfo {year} {1999})}\BibitemShut {NoStop}%
\bibitem [{\citenamefont {Cho}\ \emph {et~al.}(2005)\citenamefont {Cho}, \citenamefont {Hossain}, \citenamefont {Bae},\ and\ \citenamefont {Adesida}}]{cho2005characterization}%
  \BibitemOpen
  \bibfield  {author} {\bibinfo {author} {\bibfnamefont {H.}~\bibnamefont {Cho}}, \bibinfo {author} {\bibfnamefont {T.}~\bibnamefont {Hossain}}, \bibinfo {author} {\bibfnamefont {J.}~\bibnamefont {Bae}},\ and\ \bibinfo {author} {\bibfnamefont {I.}~\bibnamefont {Adesida}},\ }\bibfield  {title} {\enquote {\bibinfo {title} {Characterization of {Pd/Ni/Au ohmic contacts on p-GaN}},}\ }\href {https://doi.org/10.1016/j.sse.2005.01.020} {\bibfield  {journal} {\bibinfo  {journal} {Solid-State Electronics}\ }\textbf {\bibinfo {volume} {49}},\ \bibinfo {pages} {774--778} (\bibinfo {year} {2005})}\BibitemShut {NoStop}%
\bibitem [{\citenamefont {Lapeyrade}\ \emph {et~al.}(2013)\citenamefont {Lapeyrade}, \citenamefont {Muhin}, \citenamefont {Einfeldt}, \citenamefont {Zeimer}, \citenamefont {Mogilatenko}, \citenamefont {Weyers},\ and\ \citenamefont {Kneissl}}]{lapeyrade2013electrical}%
  \BibitemOpen
  \bibfield  {author} {\bibinfo {author} {\bibfnamefont {M.}~\bibnamefont {Lapeyrade}}, \bibinfo {author} {\bibfnamefont {A.}~\bibnamefont {Muhin}}, \bibinfo {author} {\bibfnamefont {S.}~\bibnamefont {Einfeldt}}, \bibinfo {author} {\bibfnamefont {U.}~\bibnamefont {Zeimer}}, \bibinfo {author} {\bibfnamefont {A.}~\bibnamefont {Mogilatenko}}, \bibinfo {author} {\bibfnamefont {M.}~\bibnamefont {Weyers}},\ and\ \bibinfo {author} {\bibfnamefont {M.}~\bibnamefont {Kneissl}},\ }\bibfield  {title} {\enquote {\bibinfo {title} {{Electrical properties and microstructure of vanadium-based contacts on ICP plasma etched n-type AlGaN:Si and GaN:Si surfaces}},}\ }\href {https://doi.org/10.1088/0268-1242/28/12/125015} {\bibfield  {journal} {\bibinfo  {journal} {Semiconductor Science and Technology}\ }\textbf {\bibinfo {volume} {28}},\ \bibinfo {pages} {125015} (\bibinfo {year} {2013})}\BibitemShut {NoStop}%
\bibitem [{\citenamefont {Haidet}\ \emph {et~al.}(2017)\citenamefont {Haidet}, \citenamefont {Sarkar}, \citenamefont {Reddy}, \citenamefont {Bryan}, \citenamefont {Bryan}, \citenamefont {Kirste}, \citenamefont {Collazo},\ and\ \citenamefont {Sitar}}]{haidet2017nonlinear}%
  \BibitemOpen
  \bibfield  {author} {\bibinfo {author} {\bibfnamefont {B.~B.}\ \bibnamefont {Haidet}}, \bibinfo {author} {\bibfnamefont {B.}~\bibnamefont {Sarkar}}, \bibinfo {author} {\bibfnamefont {P.}~\bibnamefont {Reddy}}, \bibinfo {author} {\bibfnamefont {I.}~\bibnamefont {Bryan}}, \bibinfo {author} {\bibfnamefont {Z.}~\bibnamefont {Bryan}}, \bibinfo {author} {\bibfnamefont {R.}~\bibnamefont {Kirste}}, \bibinfo {author} {\bibfnamefont {R.}~\bibnamefont {Collazo}},\ and\ \bibinfo {author} {\bibfnamefont {Z.}~\bibnamefont {Sitar}},\ }\bibfield  {title} {\enquote {\bibinfo {title} {Nonlinear analysis of {Vanadium-and Titanium-based contacts to Al-rich n-AlGaN}},}\ }\href {https://doi.org/10.7567/JJAP.56.100302} {\bibfield  {journal} {\bibinfo  {journal} {Japanese Journal of Applied Physics}\ }\textbf {\bibinfo {volume} {56}},\ \bibinfo {pages} {100302} (\bibinfo {year} {2017})}\BibitemShut {NoStop}%
\bibitem [{\citenamefont {Van~Daele}\ \emph {et~al.}(2005)\citenamefont {Van~Daele}, \citenamefont {Van~Tendeloo}, \citenamefont {Ruythooren}, \citenamefont {Derluyn}, \citenamefont {Leys},\ and\ \citenamefont {Germain}}]{van2005role}%
  \BibitemOpen
  \bibfield  {author} {\bibinfo {author} {\bibfnamefont {B.}~\bibnamefont {Van~Daele}}, \bibinfo {author} {\bibfnamefont {G.}~\bibnamefont {Van~Tendeloo}}, \bibinfo {author} {\bibfnamefont {W.}~\bibnamefont {Ruythooren}}, \bibinfo {author} {\bibfnamefont {J.}~\bibnamefont {Derluyn}}, \bibinfo {author} {\bibfnamefont {M.}~\bibnamefont {Leys}},\ and\ \bibinfo {author} {\bibfnamefont {M.}~\bibnamefont {Germain}},\ }\bibfield  {title} {\enquote {\bibinfo {title} {The role of {Al} on ohmic contact formation on n-type {GaN} and {AlGaN/ GaN}},}\ }\href {https://doi.org/10.1063/1.2008361} {\bibfield  {journal} {\bibinfo  {journal} {Applied Physics Letters}\ }\textbf {\bibinfo {volume} {87}},\ \bibinfo {pages} {061905} (\bibinfo {year} {2005})}\BibitemShut {NoStop}%
\bibitem [{\citenamefont {Kobayashi}\ \emph {et~al.}(2024)\citenamefont {Kobayashi}, \citenamefont {Sato}, \citenamefont {Okuaki}, \citenamefont {Lee}, \citenamefont {Kunimi},\ and\ \citenamefont {Kuze}}]{kobayashi2024enhanced}%
  \BibitemOpen
  \bibfield  {author} {\bibinfo {author} {\bibfnamefont {H.}~\bibnamefont {Kobayashi}}, \bibinfo {author} {\bibfnamefont {K.}~\bibnamefont {Sato}}, \bibinfo {author} {\bibfnamefont {Y.}~\bibnamefont {Okuaki}}, \bibinfo {author} {\bibfnamefont {T.}~\bibnamefont {Lee}}, \bibinfo {author} {\bibfnamefont {Y.}~\bibnamefont {Kunimi}},\ and\ \bibinfo {author} {\bibfnamefont {N.}~\bibnamefont {Kuze}},\ }\bibfield  {title} {\enquote {\bibinfo {title} {Enhanced wall-plug efficiency over 2.4\% and wavelength dependence of electrical properties at far {UV-C} light-emitting diodes on single-crystal {AlN} substrate},}\ }\href {https://doi.org/10.1002/pssr.202400002} {\bibfield  {journal} {\bibinfo  {journal} {Physica Status Solidi (RRL)--Rapid Research Letters}\ }\textbf {\bibinfo {volume} {18}},\ \bibinfo {pages} {2400002} (\bibinfo {year} {2024})}\BibitemShut {NoStop}%
\bibitem [{\citenamefont {Haidet}\ \emph {et~al.}(2015)\citenamefont {Haidet}, \citenamefont {Bryan}, \citenamefont {Reddy}, \citenamefont {Bryan}, \citenamefont {Collazo},\ and\ \citenamefont {Sitar}}]{haidet2015conduction}%
  \BibitemOpen
  \bibfield  {author} {\bibinfo {author} {\bibfnamefont {B.~B.}\ \bibnamefont {Haidet}}, \bibinfo {author} {\bibfnamefont {I.}~\bibnamefont {Bryan}}, \bibinfo {author} {\bibfnamefont {P.}~\bibnamefont {Reddy}}, \bibinfo {author} {\bibfnamefont {Z.}~\bibnamefont {Bryan}}, \bibinfo {author} {\bibfnamefont {R.}~\bibnamefont {Collazo}},\ and\ \bibinfo {author} {\bibfnamefont {Z.}~\bibnamefont {Sitar}},\ }\bibfield  {title} {\enquote {\bibinfo {title} {A conduction model for contacts to {Si-doped AlGaN grown on sapphire and single-crystalline AlN}},}\ }\href {https://doi.org/10.1063/1.4923062} {\bibfield  {journal} {\bibinfo  {journal} {Journal of Applied Physics}\ }\textbf {\bibinfo {volume} {117}},\ \bibinfo {pages} {245702} (\bibinfo {year} {2015})}\BibitemShut {NoStop}%
\bibitem [{\citenamefont {Zhang}\ \emph {et~al.}(2021)\citenamefont {Zhang}, \citenamefont {Xu}, \citenamefont {Lang}, \citenamefont {Wang}, \citenamefont {Wang}, \citenamefont {Liu}, \citenamefont {Fang}, \citenamefont {Yang}, \citenamefont {Kang}, \citenamefont {Wang}, \citenamefont {Qin}, \citenamefont {Ge},\ and\ \citenamefont {Shen}}]{zhang2021improved}%
  \BibitemOpen
  \bibfield  {author} {\bibinfo {author} {\bibfnamefont {N.}~\bibnamefont {Zhang}}, \bibinfo {author} {\bibfnamefont {F.}~\bibnamefont {Xu}}, \bibinfo {author} {\bibfnamefont {J.}~\bibnamefont {Lang}}, \bibinfo {author} {\bibfnamefont {L.}~\bibnamefont {Wang}}, \bibinfo {author} {\bibfnamefont {J.}~\bibnamefont {Wang}}, \bibinfo {author} {\bibfnamefont {B.}~\bibnamefont {Liu}}, \bibinfo {author} {\bibfnamefont {X.}~\bibnamefont {Fang}}, \bibinfo {author} {\bibfnamefont {X.}~\bibnamefont {Yang}}, \bibinfo {author} {\bibfnamefont {X.}~\bibnamefont {Kang}}, \bibinfo {author} {\bibfnamefont {X.}~\bibnamefont {Wang}}, \bibinfo {author} {\bibfnamefont {Z.}~\bibnamefont {Qin}}, \bibinfo {author} {\bibfnamefont {W.}~\bibnamefont {Ge}},\ and\ \bibinfo {author} {\bibfnamefont {B.}~\bibnamefont {Shen}},\ }\bibfield  {title} {\enquote {\bibinfo {title} {Improved ohmic contacts to plasma etched high {Al fraction n-AlGaN} by active surface pretreatment},}\ }\href {https://doi.org/10.1063/5.0042621} {\bibfield  {journal}
  {\bibinfo  {journal} {Applied Physics Letters}\ }\textbf {\bibinfo {volume} {118}},\ \bibinfo {pages} {222101} (\bibinfo {year} {2021})}\BibitemShut {NoStop}%
\bibitem [{\citenamefont {Sarkar}\ \emph {et~al.}(2017)\citenamefont {Sarkar}, \citenamefont {Haidet}, \citenamefont {Reddy}, \citenamefont {Kirste}, \citenamefont {Collazo},\ and\ \citenamefont {Sitar}}]{sarkar2017performance}%
  \BibitemOpen
  \bibfield  {author} {\bibinfo {author} {\bibfnamefont {B.}~\bibnamefont {Sarkar}}, \bibinfo {author} {\bibfnamefont {B.~B.}\ \bibnamefont {Haidet}}, \bibinfo {author} {\bibfnamefont {P.}~\bibnamefont {Reddy}}, \bibinfo {author} {\bibfnamefont {R.}~\bibnamefont {Kirste}}, \bibinfo {author} {\bibfnamefont {R.}~\bibnamefont {Collazo}},\ and\ \bibinfo {author} {\bibfnamefont {Z.}~\bibnamefont {Sitar}},\ }\bibfield  {title} {\enquote {\bibinfo {title} {Performance improvement of ohmic contacts on {Al-rich n-AlGaN grown on single crystal AlN} substrate using reactive ion etching surface treatment},}\ }\href {https://doi.org/10.7567/APEX.10.071001} {\bibfield  {journal} {\bibinfo  {journal} {Applied Physics Express}\ }\textbf {\bibinfo {volume} {10}},\ \bibinfo {pages} {071001} (\bibinfo {year} {2017})}\BibitemShut {NoStop}%
\bibitem [{\citenamefont {Ganchenkova}\ and\ \citenamefont {Nieminen}(2006)}]{ganchenkova2006nitrogen}%
  \BibitemOpen
  \bibfield  {author} {\bibinfo {author} {\bibfnamefont {M.}~\bibnamefont {Ganchenkova}}\ and\ \bibinfo {author} {\bibfnamefont {R.~M.}\ \bibnamefont {Nieminen}},\ }\bibfield  {title} {\enquote {\bibinfo {title} {Nitrogen vacancies as major point defects in gallium nitride},}\ }\href {https://doi.org/https://doi.org/10.1103/PhysRevLett.96.196402} {\bibfield  {journal} {\bibinfo  {journal} {Physical Review Letters}\ }\textbf {\bibinfo {volume} {96}},\ \bibinfo {pages} {196402} (\bibinfo {year} {2006})}\BibitemShut {NoStop}%
\bibitem [{\citenamefont {Lee}, \citenamefont {Leem},\ and\ \citenamefont {Ahn}(2000)}]{lee_annealing_2000}%
  \BibitemOpen
  \bibfield  {author} {\bibinfo {author} {\bibfnamefont {C.-R.}\ \bibnamefont {Lee}}, \bibinfo {author} {\bibfnamefont {J.-Y.}\ \bibnamefont {Leem}},\ and\ \bibinfo {author} {\bibfnamefont {B.-G.}\ \bibnamefont {Ahn}},\ }\bibfield  {title} {\enquote {\bibinfo {title} {The annealing effects of {Mg}-doped {GaN} epilayers capped with {SiO$_2$} layers},}\ }\href {https://doi.org/10.1016/S0022-0248(00)00376-6} {\bibfield  {journal} {\bibinfo  {journal} {Journal of Crystal Growth}\ }\textbf {\bibinfo {volume} {216}},\ \bibinfo {pages} {62--68} (\bibinfo {year} {2000})}\BibitemShut {NoStop}%
\bibitem [{\citenamefont {Kumabe}\ \emph {et~al.}(2024)\citenamefont {Kumabe}, \citenamefont {Yoshikawa}, \citenamefont {Kawasaki}, \citenamefont {Kushimoto}, \citenamefont {Honda}, \citenamefont {Arai}, \citenamefont {Suda},\ and\ \citenamefont {Amano}}]{kumabe2024demonstration}%
  \BibitemOpen
  \bibfield  {author} {\bibinfo {author} {\bibfnamefont {T.}~\bibnamefont {Kumabe}}, \bibinfo {author} {\bibfnamefont {A.}~\bibnamefont {Yoshikawa}}, \bibinfo {author} {\bibfnamefont {S.}~\bibnamefont {Kawasaki}}, \bibinfo {author} {\bibfnamefont {M.}~\bibnamefont {Kushimoto}}, \bibinfo {author} {\bibfnamefont {Y.}~\bibnamefont {Honda}}, \bibinfo {author} {\bibfnamefont {M.}~\bibnamefont {Arai}}, \bibinfo {author} {\bibfnamefont {J.}~\bibnamefont {Suda}},\ and\ \bibinfo {author} {\bibfnamefont {H.}~\bibnamefont {Amano}},\ }\bibfield  {title} {\enquote {\bibinfo {title} {Demonstration of {AlGaN-on-AlN} pn diodes with dopant-free distributed polarization doping},}\ }\href {https://doi.org/10.1109/TED.2024.3367314} {\bibfield  {journal} {\bibinfo  {journal} {IEEE Transactions on Electron Devices}\ }\textbf {\bibinfo {volume} {71}},\ \bibinfo {pages} {3396--3402} (\bibinfo {year} {2024})}\BibitemShut {NoStop}%
\bibitem [{\citenamefont {Schubert}(2023)}]{schubert2023light}%
  \BibitemOpen
  \bibfield  {author} {\bibinfo {author} {\bibfnamefont {E.~F.}\ \bibnamefont {Schubert}},\ }\href@noop {} {\emph {\bibinfo {title} {Light-Emitting Diodes}}}\ (\bibinfo  {publisher} {E. Fred Schubert},\ \bibinfo {year} {2023})\BibitemShut {NoStop}%
\bibitem [{\citenamefont {van Deurzen}\ \emph {et~al.}(2023)\citenamefont {van Deurzen}, \citenamefont {Singhal}, \citenamefont {Encomendero}, \citenamefont {Pieczulewski}, \citenamefont {Chang}, \citenamefont {Cho}, \citenamefont {Muller}, \citenamefont {Xing}, \citenamefont {Jena}, \citenamefont {Brandt},\ and\ \citenamefont {Lähnemann}}]{van2023excitonic}%
  \BibitemOpen
  \bibfield  {author} {\bibinfo {author} {\bibfnamefont {L.}~\bibnamefont {van Deurzen}}, \bibinfo {author} {\bibfnamefont {J.}~\bibnamefont {Singhal}}, \bibinfo {author} {\bibfnamefont {J.}~\bibnamefont {Encomendero}}, \bibinfo {author} {\bibfnamefont {N.}~\bibnamefont {Pieczulewski}}, \bibinfo {author} {\bibfnamefont {C.}~\bibnamefont {Chang}}, \bibinfo {author} {\bibfnamefont {Y.}~\bibnamefont {Cho}}, \bibinfo {author} {\bibfnamefont {D.~A.}\ \bibnamefont {Muller}}, \bibinfo {author} {\bibfnamefont {H.~G.}\ \bibnamefont {Xing}}, \bibinfo {author} {\bibfnamefont {D.}~\bibnamefont {Jena}}, \bibinfo {author} {\bibfnamefont {O.}~\bibnamefont {Brandt}},\ and\ \bibinfo {author} {\bibfnamefont {J.}~\bibnamefont {Lähnemann}},\ }\bibfield  {title} {\enquote {\bibinfo {title} {{Excitonic and deep-level emission from N-and Al-polar homoepitaxial AlN} grown by molecular beam epitaxy},}\ }\href {https://doi.org/10.1063/5.0158390} {\bibfield  {journal} {\bibinfo  {journal} {APL Materials}\ }\textbf {\bibinfo {volume}
  {11}},\ \bibinfo {pages} {081109} (\bibinfo {year} {2023})}\BibitemShut {NoStop}%
\bibitem [{\citenamefont {Yan}\ \emph {et~al.}(2025)\citenamefont {Yan}, \citenamefont {Lyons}, \citenamefont {Gordon}, \citenamefont {Janotti},\ and\ \citenamefont {Van~de Walle}}]{yan2025oxygen}%
  \BibitemOpen
  \bibfield  {author} {\bibinfo {author} {\bibfnamefont {Q.}~\bibnamefont {Yan}}, \bibinfo {author} {\bibfnamefont {J.~L.}\ \bibnamefont {Lyons}}, \bibinfo {author} {\bibfnamefont {L.}~\bibnamefont {Gordon}}, \bibinfo {author} {\bibfnamefont {A.}~\bibnamefont {Janotti}},\ and\ \bibinfo {author} {\bibfnamefont {C.~G.}\ \bibnamefont {Van~de Walle}},\ }\bibfield  {title} {\enquote {\bibinfo {title} {Oxygen impurities in {AlN} and their impact on optical absorption},}\ }\href {https://doi.org/10.1063/5.0234655} {\bibfield  {journal} {\bibinfo  {journal} {Applied Physics Letters}\ }\textbf {\bibinfo {volume} {126}},\ \bibinfo {pages} {062106} (\bibinfo {year} {2025})}\BibitemShut {NoStop}%
\end{thebibliography}%

\end{document}